\documentclass[twoside,a4paper,12pt,reqno]{amsart}
\usepackage{latexsym}
\usepackage{amsmath}
\usepackage{amssymb}
\usepackage{amsthm}
\usepackage{amsxtra}
\usepackage{amsfonts}
\usepackage{mathrsfs}
\usepackage{amsbsy}

\newcommand{\n}{\noindent}

\newcommand{\f}{\frac}
\newcommand{\ba}{\begin{eqnarray}} \newcommand{\ea}{\end{eqnarray}}
\newcommand{\be}{\begin{equation}} \newcommand{\ee}{\end{equation}}
\newcommand{\bdm}{\begin{displaymath}} \newcommand{\edm}{\end{displaymath}} 
\newcommand{\brr}{\begin{array}}\newcommand{\err}{\end{array}}

\newcommand{\bml}{\begin{gather}} 
\newcommand{\eml}{\end{gather}}

\newcommand{\vs}{\vspace{.5cm}}
\newcommand{\bo}{\boldsymbol}

\numberwithin{equation}{section}

\begin{document}

\title[ ]{
Emergence of classical trajectories in  quantum systems: 
the  cloud chamber problem in the analysis of Mott (1929) }


\author{Rodolfo Figari}

 \address{R. Figari \newline Dipartimento di Scienze Fisiche, Universit\'a di Napoli and Sezione I.N.F.N. Napoli, \newline via Cinthia, 45 - 80126 Napoli, Italy
 \newline E-mail: figari@na.infn.it}

\author{Alessandro Teta} 

\address{ A. Teta \newline   Dipartimento di Ingegneria e Scienze dell'Informazione e Matematica,   Universit\`a di L'Aquila, \newline via Vetoio (Coppito 1), 67010 L'Aquila, Italy \newline
E-mail:  teta@univaq.it}

{\maketitle}


\vs
\begin{abstract}
We analyze 
the paper 
{\em "The wave mechanics of $\alpha$-ray tracks"} (\cite{mo}), published in 1929 by N.F. Mott. In particular, we discuss  
the theoretical context in which the paper  appeared and give a detailed account of the approach 
used by the author and  the main result attained.  
Moreover, we comment on the relevance of the work not only as far as foundations of Quantum Mechanics are concerned but also as the earliest pioneering contribution in decoherence theory.
 
\end{abstract}

\vs

\section{  Introduction }
\vs

\n
In  the paper {\em "The wave mechanics of $\alpha$-ray tracks"} (\cite{mo}), published in 1929, N.F. Mott\footnote{Sir  Nevill Francis Mott (30 September 1905 - 8 August 1996) was an English physicist. He won the Nobel Prize for  Physics  in 1977 for his work on the electronic structure of magnetic and disordered systems. The award was shared with P.W. Anderson and J.H. Van Vleck (for further details see e.g. B. Pippard, Biographical Memoirs of Fellows of the Royal Society, {\bf 44}, 314-328, 1998).}  proposed a  theoretical explanation of the $\alpha$-particles  tracks observed in a Wilson cloud chamber.

\n
The work appeared in the early 
stage 
of the debate around the interpretation of Quantum Mechanics and it was 
mainly 
motivated by the attempt to clarify the meaning of the wave-particle duality and the crucial role of the measurement process. In this context the paper can be surely 
regarded as one of the seminal contributions by the 
quantum theory founders.  

\n
In addition to its foundational relevance, we consider Mott's paper extremely important 
also as the first pioneering example of the relatively recent research approach known as decoherence theory (\cite{gjkksz}). Aim of the theory is to explain the emergence of a quantum particle classical behavior as a result of its interaction with the environment. In fact, it is known that  for an isolated quantum particle an approximate classical  evolution can be derived as the limit $\hbar \rightarrow 0$ only for very special classes of initial states, like WKB or coherent states (see e.g. \cite{ro}). However, in  many important physical situations, quantum particles exhibit a quasi classical evolution for initial states drastically different from the ones in those classes.

\n
Decoherence is precisely the dynamical mechanism, entirely  predictable on the basis  of quantum evolution laws, responsible for the appearance of a particle classical behavior as a consequence of the interaction with the environment. In concrete situations the approach consists in providing models of a non trivial quantum environment which interacts with the particle going through it.

\n 
Our aim is to show that Mott's approach stands entirely inside this line of thought.  Indeed, the author analyses the $\alpha$-particle dynamics assuming a spherical wave as initial state. He attempts to explain the observed straight tracks left by the $\alpha$-particle as  a consequence of its interaction with the atoms of the vapor using a typical decoherence theory attitude.

\n
As a matter of fact, Mott's paper is less  known  than  it deserves. In the following  we  describe the context in which the work  appeared,  detail  the contents of the article  and  comment on its relevance both  in the   foundations of Quantum Theory and in the framework of decoherence theory.

\n
The paper is organized as follows.

\n
In section 2 we briefly describe the experimental and theoretical background in which
 Mott's paper appeared. In particular, we briefly recall how a Wilson chamber 
works and we  summarize the fundamental ideas leading to the
formulation of Quantum Mechanics according to the so-called Copenhagen or standard 
interpretation.

\n
In section 3 we give an account of the first theoretical arguments proposed by Born
(1927), Heisenberg (1929-30) and Darwin (1929) to explain the tracks observed in a
cloud chamber in the framework of Quantum Theory.

\n
In section 4 we present the introductory considerations appearing in Mott's paper,  describe
the  three-particle model presented by the author and give a precise formulation of the main result.

\n
In section 5 we describe in some detail Mott's arguments leading to the result and
essentially based on perturbation theory and stationary phase methods.

\n
In section 6 we conclude with some critical remarks on relevance and modernity
of Mott's paper, on some limitations of his approach and, finally, on possible
re-examinations and improvements of the result.


\vs
\section{The Wilson chamber and Quantum Theory}
\vs

\n
Roughly  speaking, a tracking chamber is a device where atoms or sub-atomic particles are detected and their very short dynamical life is recorded. 

\n
The prototype of such a device has been the cloud chamber, constructed  by Wilson in the years 1911-12 (see e.g. \cite{lr} for a description of the original apparatus). The importance of the cloud chamber  was immediately realized and the device was extensively used to investigate the properties of many different atomic and sub-atomic particles. In particular, it was used to observe and characterize the ``ionizing radiation'' emitted by radioactive sources, which was precisely  the case considered by Mott in his paper.

\n
During the last years of 19-th century Wilson was experimentally analyzing the conditions of fog formation in air saturated with water vapor. He examined the role of electric charges as condensation nuclei for the excess of vapor. In the first years of 20-th century {\em ``...ideas on the corpuscular nature of alpha- and beta-rays had become much more definite, and I had in view the possibility that the track of an ionizing particle might be made visible and photographed by condensing water on the ions which it liberated '' } (from Wilson's Nobel lecture, Stockholm, December 1927).

\n
In fact, he made available an experimental  apparatus operating schematically as follows. The air saturated with  water vapor, contained in a chamber,  is submitted to a very fast expansion lowering its temperature and driving the mixture toward a supersaturated state. The $\alpha$-particle, released  by a radioactive  source placed in the chamber, interacts with the atoms of the gas  inducing ionization. The ionized atoms  trigger the formation of small drops of water near their positions. The sequence of such drops is the visible track that one can directly observe in the chamber. Wilson had to solve the complex experimental problem of instantaneous and synchronized illumination and photography of the sequence of drops.

\n
 The track is  considered as the experimental manifestation of the "trajectory" of the $\alpha$-particle and, as a matter of fact,  is accurately described as the trajectory of a classical particle (relativistic or non relativistic according to the initial particle velocity)   in a classical electromagnetic field. In particular, one observes straight lines whenever no electromagnetic field is present.

\n
The advent of Quantum Mechanics  in the years 1925-27, together with its standard interpretation,   made the effectiveness of the classical description of the tracks rather problematic. 

\n
In order to recall the context at the time the cloud chamber problem was first approached we briefly summarize here some crucial steps which led to the final formulation of the theory (see e.g. \cite{ja}, \cite{c}, \cite{ck}, \cite{st})

\n
As it is well known,  the decisive contributions to  the elaboration  of the theory came from Heisenberg (\cite{he1}), together with Born and Jordan (\cite{bj}) in 1925, and from Schr\"odinger (\cite{sc}) in 1926. 
The two approaches, known as Matrix Mechanics and Wave Mechanics respectively,  were based on radically different  physical descriptions of atomic phenomena. 

\n
The former approach is characterized by the explicit rejection of the classical  notions of position and velocity in the description of atomic phenomena, in favor of a new kinematics based on observable quantities, like frequencies and amplitudes of the emitted or absorbed radiation in quantum jumps. As a consequence, Matrix Mechanics appears as an abstract algebraic formalism where any visualized or intuitive description of a microscopic object is abandoned.

\n
On the contrary,  according to Schr\"odinger, such intuitive space-time description is still possible if one admits that a microscopic object is  described by a wave instead of a point particle. On the basis of the analogy between optics and mechanics, he was able to derive the evolution equation for the wave $\psi(x,t)$. Moreover, he also proposed a first physical interpretation  of $e |\psi(x,t)|^2$ as the density charge at the point $x$ and time $t$ of the microscopic object with total charge $e$.

\n
The mathematical equivalence of the two theories was soon proved by Schr\"odinger himself leaving open the problem to provide the correct  physical interpretation of the formal structure. 

\n
The Schr\"odinger proposal encountered severe difficulties due to the fact that, in general, the solutions of the evolution equation inevitably spread in space as time grows. The finally accepted interpretation was given by Born (\cite{bor}) in 1926. According to his proposal,  $|\psi(x,t)|^2$ is the probability density to find the object in $x$ at time $t$. More generally, after Born,  Quantum Mechanics was accepted as a theory which can only provide probabilistic predictions (of the position or  of any other observable relative to a microscopic object). 

\n
A further important step in the direction of a deeper understanding of the formalism  was made  in 1927 by Heisenberg. In  his paper on the uncertainty principle (\cite{he2})  he was able to prove the uncertainty relations for position and momentum as a consequence of the non commutativity of the corresponding operators $q$ and $p$. The physical implications of the uncertainty relations were illustrated through the famous thought experiment with a gamma-ray microscope to determine the exact position of an electron. The conclusion was that a more accurate determination  of the position of the electron implies  a less accurate determination of the momentum and vice versa. According to Heisenberg, the operational  impossibility to determine both position and momentum of a particle was the origin of the statistical nature  of Quantum theory, making, at the same time,  the classical notions of position, momentum and trajectory inadequate to describe the microscopic world.

\n
The work of Heisenberg on uncertainty relations stimulated  Bohr to   clarify his point of view on the interpretation of the theory.  The main purpose of Bohr  was to harmonize Wave and Matrix Mechanics in a unified and coherent description of atomic phenomena.    The occasion was the lecture delivered at the congress in Como, on September 1927 (\cite{boh}), and also the discussions in the subsequent Solvay Conference in Brussels, on October 1927 (\cite{bv}). 
Bohr's approach soon became the core of the so-called Copenhagen or standard interpretation of Quantum Mechanics.

\n
The main features of such interpretation scheme can be roughly summarized as follows: i) completeness, ii) wave-particle duality, iii) lack of causal and space-time description, iv) crucial role of the observation.

\n
i) Completeness means that the wave function $\psi $ describes the actual state of an individual system and the probabilistic meaning of $|\psi|^2$ has an ontological character, i.e. it does not refer to our ignorance of hidden parameters. 

\n
ii) Wave-particle duality should be understood in the sense that a microscopic system cannot be described in terms of only one of the two classical concepts of wave or particle. The system rather behaves either as a wave or as a particle, depending on the context or, more precisely, on the experimental apparatus used to observe the system.  Therefore wave and particle  behavior are two mutually exclusive and complementary aspects  of the microscopic object essence.

\n
iii) According to Bohr, Heisenberg uncertainty relations show the impossibility of having a causal description (based on the precise determination of the momentum) and a space-time description (based on the precise determination of the position) simultaneously. In this sense also causal and space-time descriptions are classical concepts which must be considered as two mutually exclusive and complementary aspects of the description of a quantum system.

\n
iv) The last important point  is the role of the observation. A measurement apparatus must be considered as a classical object,  characterized by a precise determination of its classical properties. The measurement process consists in the interaction between the microscopic system and the classical apparatus. The result of a measurement is the determination of one of the possible complementary properties of the quantum system. The crucial point is that the measured property of the system cannot be thought as pre-existing to the measurement process.  The property is produced only as the result of the interaction with the classical apparatus. More precisely, the emerging property is determined by the whole experimental context, i.e. the specific preparation of the system and the characteristics of the apparatus. 
This, in particular, means that the  state of the system immediately before the measurement (when the system does not possess the given property)    is different from the state immediately after the measurement (when the system possesses the property). The abrupt change of the state of the system determined by  the measurement  process is called wave packet reduction.

\n
It is worth mentioning that the above conception of the measurement process is the most delicate aspect of the Copenhagen interpretation and it has raised a long debate that still continues. Here we only briefly mention some problematic points. 

\n
First,  it is not explained why all or part of the experimental device, despite being made of atoms, must behave as a classical object with well-defined classical properties.

\n
Moreover it is not clear where the borderline between the measurement apparatus (characterized by a classical behavior) and the system (characterized by a  quantum behavior) should be put. The problem is usually solved pragmatically in each specific situation but, at a conceptual level, the  ambiguity remains. 

\n
Finally, one has to  renounce  to describe  the interaction of the system with the apparatus using the Schr\"odinger equation. This fact was formalized in 1932  by von Neumann (\cite{vo}) who   postulated two different kind of evolution for the system: a genuine quantum evolution governed by the Schr\"odinger equation when the system is not measured and a different (stochastic and non linear) evolution producing the wave packet reduction when the system is measured.

\n
We only mention that many attempts have been made to clarify  these conceptual problems within the framework of the standard interpretation but a reasonable and universally accepted solution has not been found. 

\n
From the point of view of our analysis, it is interesting to observe that the problematic aspects of the system-apparatus interaction are immediately evident when one approaches the description of a quantum particle in a cloud chamber.

\n
In fact, for a quantum mechanical description of the process one must take into account  that the initial state of the particle emitted by the source does not have the form of a semi-classical wave packet but rather a highly correlated continuous superposition of states with well localized position and momentum. 
In particular, according to the first theoretical analysis of the radioactive decay given by Gamow (\cite{ga})
, the emitted $\alpha$-particle must be described by a wave function having the form of a spherical wave, with center in the radioactive nucleus and isotropically propagating in space.  Therefore, the non trivial problem arises as to how such a superposition state can produce the observed classical trajectories. As we shall see in the next sections, in such explanatory scheme a crucial role is played by the act of  "observation",  where the system to be measured can be the $\alpha$-particle (in particular its position) as well as the atoms of the vapor (in particular their excitation).

\vs
\vs

\section{Early contributions }
\vs

\n
In this section we shall briefly summarize the first theoretical attempts  to explain the apparent contradiction between the highly correlated superposition state of the $\alpha$-particle and the observed tracks in the chamber. In particular, we shall consider the contributions of Born (1927), Heisenberg (1929-30) and Darwin (1929).

\vspace{0.3cm}
\n
3.1 Born (1927). 

\n
The  theoretical explanation of the observed tracks in a cloud chamber was already approached by Born (for the first time to our knowledge) in 1927 during the general discussion at the Solvay Conference (\cite{bv}). In his words: {\em "Mr. Einstein has considered the following problem: A radioactive sample emits $\alpha$-particles in all directions; these are made visible by the method of the Wilson cloud chamber. Now, if one associates a spherical wave with each emission process, how can one understand that the track of each $\alpha$-particle appears as a (very nearly) straight line? In other words: how can the corpuscular character of the phenomenon be reconciled here with the representation by waves?"}

\n
According to Born, the answer can be given resorting to the notion of {\em "reduction of the probability packet"} discussed by Heisenberg in \cite{he3}.  According to this notion,  the observation of the electron position by light of wavelength $\lambda$ would produce the reduction of the probability packet  for the position of the electron to a region of linear size $\lambda$. In the subsequent evolution the packet spreads in space until a new observation reduces it  again to a packet of width $\lambda$. Then, in Heisenberg's words: {\em "Every determination of position reduces therefore the wave packet back to its original size $\lambda$"}. This mechanism of reduction would be responsible for the appearance of a (nearly) classical trajectory of the electron. 

\n
The same idea is used by  Born in the case of the cloud chamber. Here the observation by means of light must be replaced by ionization of the atoms of the vapor in the chamber: {\em "As soon as such ionization is shown by the appearance of cloud droplets, in order to describe what happens afterwards one must reduce the wave packet in the immediate vicinity of the drops. One thus obtains a wave packet in the form of a ray, which corresponds to the corpuscular character of the phenomenon"}. 

\n
It is worth to emphasize that, according to this reasoning, the whole process is  described in terms of the interaction of a quantum system (the $\alpha$-particle) with a classical measurement apparatus (the atoms of the vapor).  Such interaction, which is not described by Schr\"odinger equation, produces  "reduction" of the spherical wave to a wave packet with  definite position and momentum.

\n
Following a suggestion by Pauli, Born continues discussing the possibility to describe  both the $\alpha$-particle and the atoms of the vapor as constituents of a unique quantum system, whose wave function depends on the coordinates of all the particles of the system. 

\n
In particular, Born proposes, as an example, a simplified one dimensional model consisting of the $\alpha$-particle and only two atoms. The $\alpha$-particle is initially in a superposition   state of two wave packets with opposite momentum and position close to the origin. The two cases where atoms are placed on the same side or on opposite sides with respect to the origin are considered. Born  discusses, at a purely qualitatively level, the probability that during the time evolution of the whole system the two atoms will be hit in the two different cases. We simply give the conclusions reached by Born without going into details (for a quantitative analysis of the same model  see e.g. \cite{dft1}).

\n
The first statement he  makes, without invoking the reduction of the wave packet,  is that the $\alpha$-particle has negligible probability to hit both atoms unless they are on the same side with respect to  the origin.

\n
On the other hand, Born concludes  his argument having an explicit recourse to the reduction postulate, saying: {\em "To the reduction of the wave packets corresponds the choice of one of the two directions of propagations}, and the choice is made as soon as one observes the excitation of an atom, as a consequence of the collision. Only starting from such observation the evolution of the $\alpha$-particle can be described as a real classical trajectory.

\n
In conclusion, it seems that according to Born a more detailed description of the $\alpha$-particle in a cloud chamber taking into account the presence of the environment is in fact possible {\em "but this does not lead us further as regards the fundamental questions."}

\vspace{0.3cm}
\n
3.2 Heisenberg (1929-30). 

\n
As already mentioned, Born's analysis was explicitly inspired by the considerations of Heisenberg in \cite{he2} on the applicability of the notion of classical trajectory of an electron. It is remarkable that  Heisenberg himself explicitly reconsidered   the cloud chamber problem in his lectures  at the University of Chicago in 1929, published in \cite{he3}, following the same line of thought   of Born but with considerable more details.

\n

\n
His approach, described in chapter 5 of the book,  can be considered an exhaustive qualitative investigation of the problem according to the standard interpretation of Quantum Mechanics. This analysis  had a deep influence on the physics community for many years.

\n
Heisenberg's  first (rather extreme) remark is that the whole experimental situation could be satisfactorily described using only Classical Mechanics, but  it might also be of interest to discuss the problem from the point of view of Quantum Theory.

\n
He stresses that, approaching a quantum theoretical description, one is immediately faced  with the problem of separating the quantum system from the apparatus. In the case of the cloud chamber, one has two different reasonable choices: a) the quantum system consists of the $\alpha$-particle alone (and then the molecules of the vapor play the role of the measurement apparatus); b) the quantum system consists of the $\alpha$-particle and the molecules of the vapor.

\n
It should be emphasized that the physical descriptions of the experimental situation obtained following the  two different choices are explicitly considered equivalent.

\n
Heisenberg's line of reasoning in examining the two cases proceeds as follows.

\n
In case a) the single molecule of the vapor measures the position of the $\alpha$-particle. Assume that  the molecule (supposed at rest)  occupies a volume $\Delta q$ around the point $q_1$ and  $t_1$ is the collision time between $\alpha$-particle and molecule. As result of the measurement process, the state of the $\alpha$-particle is suddenly reduced and therefore  at time $t_1$ it has position $q_1$ with spread $\Delta q$. On the other  hand one knows the position $q_0$ of the $\alpha$-particle at time $t_0$, when it leaves the radioactive source. Since no external force is present, one   infers that the momentum of the $\alpha$-particle at time $t_1$ is $p_1= M (q_1-q_0)/(t_1 -t_0)$, where $M$ denotes the mass.

\n
It is then possible  to conclude that for $t> t_1$ the $\alpha$-particle is described by the free evolution of a wave packet starting from $q_1$, with initial spread $\Delta q$ and momentum along the straight line $\gamma$ joining $q_0$ and $q_1$. Therefore, the center of the wave packet moves along the same line $\gamma$.

\n
During the time evolution the wave packet inevitably spreads out. On the other hand the $\alpha$-particle  collides with other molecules placed along $\gamma$ and after each collision the same measurement process of the position takes place. In this way the spreading is repeatedly reduced and the wave packet remains focused around the straight line $\gamma$, corresponding to the observed "trajectory" of the $\alpha$-particle.

\n
Let us consider case b), where the molecules of the vapor are considered part of the quantum system. Heisenberg starts with an interesting claim that  would have probably deserved  further analysis:   in case b) the procedure to account for the the observed trajectories is more complicated but, on the other hand, it allows to hide the role of the reduction of the wave packet.

\n
Then he goes on to describe a simplified model made of the $\alpha$-particle plus only two molecules. The molecules are non interacting, their centers of mass are fixed in the positions $a_I$, $a_{II}$ and the internal coordinates are denoted by $q_I$, $q_{II}$. It is assumed that the Hamiltonians of the two molecules have a complete set of eigenfunctions  $\varphi_{n_I}(q_I)$, $\varphi_{n_{II}} (q_{II})$, corresponding to a discrete set of eigenvalues labeled by  integers $n_I$, $n_{II}$. 

\n
The initial state of the system is chosen in the form of a  product of the ground states of the molecules  (labeled by $n^0_I$, $n^0_{II}$)  times a plane wave with  momentum $p$ for the  $\alpha$-particle.

\n
The interesting object to compute is the probability that both molecules are excited and the result of the computation is that such a probability is significantly different from zero only if the momentum $p$ is parallel to the line joining $a_I$ and $a_{II}$. Since the passage of the $\alpha$-particle is indirectly observed through  the excitations of the molecules, the result explains why one can only see straight trajectories in a cloud chamber.

\n
The solution of the Schr\"odinger equation for the three-particle system  is approached treating  the interaction between the $\alpha$-particle and the molecules as small perturbation of the free dynamics and assuming that the momentum $p$ is large with respect to the spacing of energy levels of the molecules.

\n
The perturbative computation is not developed  in all details by Heisenberg. Here we only summarize the main steps of his procedure in order to clarify the line of reasoning.

\n
The Schr\"odinger equation is solved by iteration and at the first order the wave function  can be written in the form
\be
\psi^{(1)}= e^{-i\f{t}{\hbar} E_0} \sum_{n_I , n_{II}} w^{(1)}_{n_I n_{II}} (x) \varphi_{n_I}(q_I)\varphi_{n_{II}} (q_{II})
\ee
where $E_0$ denotes the total energy of the system, $x$ the position coordinate of the $\alpha$-particle and the coefficients $ w^{(1)}_{n_I n_{II}} $ satisfy an equation which is easily derived from the original Schr\"odinger equation. 
From Born's rule it follows that $|w^{(1)}_{n_I n_{II}} (x)|^2$ represents the probability density (at first order) to find the $\alpha$-particle in $x$  when the molecules are in the states labeled by $n_I$, $n_{II}$.

\n
From the equation for $ w^{(1)}_{n_I n_{II}} $ it is immediately evident the first result:  the probability that both molecules are excited is zero at first order.

\n
The second result claimed by Heisenberg is definitively less evident and nevertheless it is stated without a detailed proof. It says that  $ w^{(1)}_{n_I n^0_{II}} $ is significantly different from zero only in a strip parallel to $p$ located behind the molecule $I$, whose thickness (close to the molecule) is of the same order of the dimension of the molecule. The same kind of result is obviously true for $ w^{(1)}_{n^0_I n_{II}}$.

\n
Then he continues with the analysis of the wave function at the second order
\be
\psi^{(2)}=e^{-i\f{t}{\hbar} E_0} \sum_{n_I , n_{II}} w^{(2)}_{n_I n_{II}} (x) \varphi_{n_I}(q_I)\varphi_{n_{II}} (q_{II})
\ee
Writing the equation for $ w^{(2)}_{n_I n_{II}}$ and exploiting the results obtained at the first order, he derives the desired final result.  The probability density at second order $| w^{(2)}_{n_I n_{II}} (x)|^2$, with $n_I \neq n^0_I$ and $n_{II} \neq n^0_{II}$,  is significantly different from zero only if the following  two situations occurs: the molecule $II$ is in the strip of $w^{(1)}_{n_I n^0_{II}}$ or the molecule $I$ is in the strip of $w^{(1)}_{n^0_I n_{II}}$.

\n
The procedure can be iterated with an arbitrary number of molecules and therefore the linearity of the trajectories is proved.

\n
At the end of the analysis, Heisenberg makes a second claim on the problem of the wave packet reduction. In particular, he explains that in case b) the reduction  takes place when one arranges a measurement process "to observe" the excitation of the molecules. This simply means that  the unavoidable line of separation between the system and the apparatus has been  moved  to include the molecules in the system.  
In this sense one  should  probably understand the previous claim that the reduction in case b) is hidden.

\n
Summarizing,  Heisenberg's analysis of the cloud chamber, like Born's analysis, insists to consider  the treatments in case a) and b) as conceptually equivalent. This belief is founded on the fact that in any case  the recourse to the reduction of the wave packet cannot be avoided.

\vspace{0.3cm}
\n
3.3 Darwin (1929).

\n
A further relevant contribution to the explanation of the trajectories in a cloud chamber was given by Darwin  (\cite{da})  in 1929. In such paper 
 one does not find any analysis of specific models. Nevertheless  there is a detailed discussion of the problem and it is clearly stated a possible strategy for an approach entirely based on the use of  Schr\"odinger equation.

\n
Darwin approaches a collision problem in the framework of Wave Mechanics with the aim to {\em "take a problem which would be regarded at first sight as irreconcilable with a pure wave theory, but thoroughly typical of the behavior of particles, and show how in fact the correct result arises naturally from the consideration of waves alone."}

\n
He emphasizes that in order to obtain the correct predictions on the behavior of a given system $\mathcal{S}$ one has to take into account its interaction with (part of) the environment $\mathcal{E}$. Therefore the wave function $\psi$ is not a wave in ordinary three dimensional space but rather it is a function  of the coordinates of $\mathcal{S}$ and of $\mathcal{E}$. Only when such $\psi$ has been computed, the probabilistic predictions on $\mathcal{S}$ are obtained by taking an average over all possible final configurations of $\mathcal{E}$.

\n 
Such procedure, even if  {\em "discouragingly complicated"}, can account for the particle-like behavior working only on $\psi$ and  without invoking any act of observation. It is worth noticing that the program outlined by Darwin essentially coincides with the basic strategy of modern decoherence theory.

\n
After these considerations, Darwin discusses  a concrete example where the intuitive particle behavior can be derived from the analysis of the wave function.  That part of the analysis is not directly connected with the cloud chamber and therefore it is not relevant for our purposes. 
However, in the final part of the paper, one finds some other interesting considerations. 

\n
He examines the case of the ray tracks of $\alpha$-particles in a cloud chamber, {\em "one of the most striking manifestations of particle characters"},  in connection with Gamow's theory of radioactive decay (\cite{ga}), distinguishing two different points of view. 

\n
According to the first one  {\em "we must regard Gamow's calculations as determining only the probability of disintegration, and that when this has taken place, we start the next stage by assigning a definite direction for the motion of the $\alpha$-particle; after which we reconvert it into a wave, but now on a narrow front, so as to find its subsequent history"}. 

\n
As an alternative to this point of view, he first notices that $\alpha$-rays can in principle exhibit diffraction and therefore it is reasonable to assign a real existence to the spherical wave outside the nucleus. Then he  discusses a possible wave description of the experiment. The wave function $\psi$ is a function of the coordinates of the $\alpha$-particle and of the coordinates of the atoms in the chamber and, before the first collision, it is a product of the spherical wave for the $\alpha$-particle times a set of stationary (in general ground) states for the atoms. {\em "But the first collision changes this product into a function in which the two types of coordinates are inextricably mixed, and every subsequent collision makes it worse."}   Such complicated function contains a phase factor and {\em "without in the least seeing the details, it looks quite natural to expect that this phase factor will have some special character, such as vanishing, when the various co-ordinates  satisfy a condition of collinearity."} 

\n
It is interesting to notice that Darwin clearly identifies stationary phase arguments as the crucial technical tools required to predict the particle-like behavior.

\n
Then he continues: {\em ``So without pretending to have mastered the details, we can understand how it is possible that the $\psi$ function, so to speak, not to know in what direction  the track is to be, but yet to insist that it should be a straight line. The decision as to actual track can be postponed until the wave reaches the uncovered part, where the observations are made.''}

\n
This approach seems to have a general validity. In Darwin's view the  wave-particle duality proposed by Bohr can be avoided. The whole quantum theory can be based on the wave function $\psi$, considered as the central object from which all the particle or wave properties can be accurately described, at least until a real measurement is performed.  In his words  {\em "it thus seems legitimate to suppose that it is always admissible to postpone the stage, at which we are forced to think of particles, right up to the point at which they are actually observed."}


\vs

\section{Mott's paper}

\vs
\n
The program enunciated by Darwin was concretely realized by  Mott in his  seminal paper of 1929. 

\n
In the introduction Mott recognizes to have been inspired by Darwin's paper in his attempt to explain the typical particle-like properties of an $\alpha$-particle in a cloud chamber using only Wave Mechanics. He admits that such point of view seems at first sight counterintuitive , since  {\em "it is a little difficult to picture how it is that an outgoing spherical wave can produce a straight track; we think intuitively that it should ionise atoms at random throughout space"}. 
Like Heisenberg, Mott points out that  the crucial point is to establish the borderline between the system under consideration  and the measuring device.  
He recalls the two possible approaches: in the first the $\alpha$-particle is the quantum system under consideration (and the gas of the chamber is part of the measuring device) while in the second approach the quantum system consists of the $\alpha$-particle and of the atoms of the gas.

\n
Mott proceeds toward a detailed analysis of the problem following closely the latter approach. He claims that the intuitive difficulty mentioned above  can be overcome since it arises from our erroneous {\em "tendency to picture the wave as existing in ordinary three dimensional space, whereas we are really dealing with wave functions in multispace formed by the co-ordinates both of the $\alpha$-particle and of every atom in the Wilson chamber".}  

\n
The model considered by Mott consists of the $\alpha$-particle, initially described by a spherical wave centered at the origin, and  the electrons in two hydrogen atoms. The nuclei  of the atoms are supposed at rest in the fixed positions $\bf{a_1}$, $\bf{a_2}$, with $|\bo a_1| < |\bo a_2|$. It is assumed that  the $\alpha$-particle does not interacts with the nuclei and that the interaction between the two electrons is negligible. The interaction between the $\alpha$-particle and the electrons is assumed to be weak. 

\n
The model is essentially the same considered by Heisenberg in case b). Nevertheless, as we shall see,   Mott's analysis is definitely more detailed and therefore the outcome appears more transparent and convincing. 

\n
The main result of the paper can be summarized in the following statement:

\n
under suitable assumptions (which will be specified later) the two hydrogen atoms cannot both  be excited (or ionized) unless $\bf{a_1}$, $\bf{a_2}$ and the origin lie on the same straight line. 

\n
We shall describe how Mott derives the result trying to follow the original notation and line of reasoning.

\n
Main objects of the  investigation are periodic solutions $F(\bo{R},\bo{r_1},\bo{r_2})  e^{iEt/\hbar}$ of the Schr\"odinger equation for the three particle system, where  $\bo{R}$, $\bo{r_1}$, $\bo{r_2}$ denote the coordinates of the $\alpha$-particle and of the two hydrogen atom electrons respectively. The function  $F$ (depending parametrically on $E$)  is  solution of the stationary Schr\"odinger equation
\ba\label{eqF}
&&-\f{\hbar^2}{2M} \Delta_R F +\Big(  -\f{\hbar^2}{2m} \Delta_{r_1}     -\f{e^2}{|\bo{r_1}-\bo{a_1}|} \Big) F + \Big(
-\f{\hbar^2}{2m} \Delta_{r_2}    -\f{e^2}{|\bo{r_2}-\bo{a_2}|} \Big) F \nonumber\\ 
&&- \Big( \f{2e^2}{|\bo{R}-\bo{r_1}| }  + \f{2e^2}{|\bo{R}-\bo{r_2}| } \Big) F
 =E\, F
 \ea
where $\Delta_x$ is the laplacian with respect to the coordinate $x$, $M$ is the mass of the $\alpha$-particle,  $m$ is the mass of the electron,  $-e$ is the charge of the electron so that  $2e$ is the charge of the $\alpha$-particle. 

\n
The solution of equation \eqref{eqF} can be conveniently expanded in series of  eigenfunctions of the two electrons in the hydrogen atoms. More precisely, let $\psi_j$ be the $j$-th eigenfunction of an hydrogen atom centered in the origin, with $\psi_0$ denoting the ground state. Then the corresponding eigenfunctions of the atoms in $\bf{a_1}$, $\bf{a_2}$ are
\be
\Psi^I_j (\bo{r_1}) = \psi_j(\bo{r_1}-\bo{a_1}), \;\;\;\;\;\;\Psi^{II}_j (\bo{r_2}) = \psi_j(\bo{r_2}-\bo{a_2})
\ee
We notice that here it is tacitly assumed that the index $j$ can be an integer or a real positive number (and correspondingly $\psi_j$ can be  a proper eigenfunction or a generalized eigenfunction). 

\n
Exploiting completeness of the system of  eigenfunctions, we have the following representation for $F$

\be\label{espF}
F(\bo{R},\bo{r_1},\bo{r_2}) = \sum_{j_1, j_2} f_{j_1 j_2}(\bo{R}) \Psi^I_{j_1} (\bo{r_1}) \Psi^{II}_{j_2} (\bo{r_2}) 
\ee
The Fourier coefficients $f_{j_1 j_2}(\bf{R})$ of the expansion have a direct physical interpretation. Indeed, exploiting Born's rule, the probability to find the first atom in the state labeled by $j_1$ and the second atom in the state labeled by $j_2$ is 
\be\label{prob}
\int \!\! d\bo{R}\, |f_{j_1 j_2}(\bo{R})|^2
\ee
One can pictorially say that the "wave function"  of  the $\alpha$-particle is   $f_{00}(\bf{R})$ if both atoms remain in the ground state,  $f_{j_1 0}(\bf{R})$, $j_1 \neq 0$,  if the first atom is in the $j_1$-th excited (or ionized) state and the second in the ground state,  $f_{j_1 j_2}(\bf{R})$, $j_1, j_2 \neq 0$, if both atoms  are excited (or ionized). 

\n
The  analysis will show   that  $f_{00}(\bf{R})$ is a (slightly deformed) spherical wave and $f_{j_1 0}(\bf{R})$, $j_1 \neq 0$, is a wave packet emerging from $\bf{a_1}$ with a momentum along the line $\overline{O\bf{a_1}}$. This means that the second atom can be excited  by such wave packet only if $\bf{a_2}$ lies on the line $\overline{O\bf{a_1}}$. Thus  the desired result will follow, i.e.  $f_{j_1 j_2}(\bf{R})$, $j_1, j_2 \neq 0$, is approximately zero unless the condition of co-linearity is satisfied.


\vs
\section{Derivation of the result}
\vs

\n
Exploiting the assumed weakness of the interaction between $\alpha$-particle and electrons, 
the computation is carried out using second order perturbation theory. If we write
\be
F=F^{(0)} + F^{(1)}+F^{(2)} + \cdots
\ee 
then by  successive approximation one has for $n\geq 1$
\ba \label{asF}
&&-\f{\hbar^2}{2M} \Delta_R F^{(n)} +\Big(  -\f{\hbar^2}{2m} \Delta_{r_1}     -\f{e^2}{|\bf{r_1}-\bf{a_1}|} \Big) F^{(n)} + \Big(
-\f{\hbar^2}{2m} \Delta_{r_2}    -\f{e^2}{|\bf{r_2}-\bf{a_2}|} \Big) F^{(n)} - E \,F^{(n)} \nonumber\\ 
&&= \Big( \f{2e^2}{|\bf{R}-\bf{r_1}| }  + \f{2e^2}{|\bf{R}-\bf{r_2}| } \Big) F^{(n-1)}
\ea
\label{0 ord} For $n=0$ one has to solve the unperturbed equation
\be\label{0F}
-\f{\hbar^2}{2M} \Delta_R F^{(0)} +\Big(  -\f{\hbar^2}{2m} \Delta_{r_1}     -\f{e^2}{|\bf{r_1}-\bf{a_1}|} \Big) F^{(0)} + \Big(
-\f{\hbar^2}{2m} \Delta_{r_2}    -\f{e^2}{|\bf{r_2}-\bf{a_2}|} \Big) F^{(0)} - E \, F^{(0)}=0 
\ee
As solution of \eqref{0F} one chooses a diverging spherical wave times the ground state of the two atoms
\be\label{00F}
F^{(0)}(\bo{R}, \bo{r_1}, \bo{r_2})\!= f_{00}^{(0)}(\bo R)  \Psi_0^{I}(\bo{r_1}) \Psi_0^{II} (\bo{r_2})\,, \;\;\;\;\;\; f_{00}^{(0)}(\bo R) =  \f{e^{i k |\bo{R}|}}{|\bo{R}|} \,, \;\;\;\;\;\; k=\!\f{\sqrt{2M(E\!-\!2E_0)}}{\hbar}
\ee
where $E_j$ denotes the $j$-th eigenvalue of the hydrogen atom and $f_{00}^{(0)}$ is the outgoing solution of the Helmotz equation. Notice that the use of stationary Schr\"odinger equation forces Mott to choose a solution not in $L^2$ which, strictly speaking, is not legitimate. In particular, the probabilistic interpretation \eqref{prob} fails for  $f_{00}^{(0)}$. 

\n
For the first order term $F^{(1)}$ we write
\be\label{exF1}
F^{(1)} (\bo{R},\bo{r_1},\bo{r_2}) = \sum_{i_1, i_2} f^{(1)}_{i_1 i_2}(\bo{R}) \Psi^I_{i_1} (\bo{r_1}) \Psi^{II}_{i_2} (\bo{r_2}) 
\ee
Substituting \eqref{exF1} into \eqref{asF} for $n=1$ one has
\ba  \label{eqf1}
&&\sum_{i_1,i_2} \Big(\!\! -\f{\hbar^2}{2M} \Delta_R +E_{i_1} + E_{i_2} -E \Big)  f^{(1)}_{i_1 i_2}(\bo{R}) \Psi^I_{i_1} (\bo{r_1}) \Psi^{II}_{i_2} (\bo{r_2}) \nonumber \\ 
&&=  \Big( \f{2e^2}{|\bf{R}-\bf{r_1}| }  + \f{2e^2}{|\bf{R}-\bf{r_2}| } \Big)   f_{00}^{(0)}(\bo R)  \Psi_0^{I}(\bo{r_1}) \Psi_0^{II} (\bo{r_2})
\ea
Multiplying the above equation by $\Psi^I_{j_1} (\bo{r_1}) \Psi^{II}_{j_2} (\bo{r_2})$  and integrating over the coordinates of the electrons one obtains the equation for $ f^{(1)}_{j_1 j_2}(\bo{R})$
\ba\label{eqf12}
&& \Big(\!\! -\f{\hbar^2}{2M} \Delta_R +E_{j_1} + E_{j_2} -E \Big)  f^{(1)}_{j_1 j_2}(\bo{R}) =   f_{00}^{(0)}(\bo R) \Big(  \delta_{j_2 0} \! \int\!\! d \bo r_1  \,   \f{2e^2}{|\bf{R}-\bf{r_1}| } \Psi_0^{I}(\bo{r_1}) \Psi^I_{j_1} (\bo{r_1})  
\nonumber\\
&& + \, \delta_{j_1 0} \! \int\!\! d \bo r_2  \,   \f{2e^2}{|\bf{R}-\bf{r_2}| } \Psi_0^{II}(\bo{r_2}) \Psi^{II}_{j_2} (\bo{r_2})  \Big)
\ea
Equation \eqref{eqf12} can be rewritten in the more compact form
\be\label{eqf13}
 \Big( \f{\hbar^2}{2M} \Delta_R +E -E_{j_1} - E_{j_2} \Big)  f^{(1)}_{j_1 j_2}(\bo{R}) =
K_{j_1 j_2}(\bo R)
\ee
where
\be\label{K}
K_{j_1 j_2}(\bo R)=  f_{00}^{(0)}(\bo R) \Big(  \delta_{0 j_2 } V_{ j_1 0} (\bo R - \bo a_1) + \delta_{j_1 0} V_{0 j_2} (\bo R - \bo a_2) \Big)
\ee
\be\label{V0j}
V_{i j} (\bo x) = - \int\!\! d \bo y \, \f{2 e^2}{|\bo x - \bo y|} \psi_i(\bo y) \psi_j (\bo y)
\ee

\n
The computation of the first order term $F^{(1)}$ is now reduced to the solution of the non-homogeneous Helmotz equation  \eqref{eqf13}.  Mott refers to the treatise by Courant and Hilbert, Methoden der Mathematischen Physik, chap. 5, par. 10, to assert that the most general solution is
\be\label{solf1}
f_{j_1 j_2}^{(1)}(\bo R)= G_{j_1 j_2}(\bo R) + \f{M}{2 \pi \hbar^2} \int\!\! d\bo R' \,K_{j_1 j_2} (\bo R')\, \f{e^{\pm i k'|\bo R - \bo R'|}}{|\bo R - \bo R'|} \,,\;\;\;\;\;\;\;\; k'=\f{\sqrt{2M(E-E_{j_1} - E_{j_2})}}{\hbar}
\ee
where $G_{j_1 j_2}$ is an arbitrary solution of the homogeneous equation $(\Delta +k'^2)G_{j_1 j_2}=0$.  In our case the phase in the exponential must be taken with the sign $+$ since we are interested in waves diverging from $\bo a_1$ or $\bo a_2$. 
  Moreover Mott argues that $G_{j_1 j_2}$ "{\em represents streams of particles fired at already excited atoms"} while, as  initial condition, we have both atoms in their ground state. Therefore one must require
\be\label{G}
G_{j_1 j_2}=0
\ee
Furthermore, from \eqref{K} one sees that $K_{j_1 j_2}(\bo R) =0$ if both $j_1$ and $j_2$ are different from zero and therefore, by \eqref{solf1}, one also has $f_{j_1 j_2}^{(1)} =0$ if both $j_1$ and $j_2$ are different from zero. 

\n
From these preliminary considerations a first  conclusion can be drawn:

\n
at first order in perturbation theory the probability that both atoms are excited is always zero.

\n
The result is not surprising since, as Mott remarks, in perturbation theory the probability that one atom is excited is a first order quantity while the probability that both atoms are excited is a second order quantity. This explains why the second order term $F^{(2)}$ is required in order to obtain an estimate of a double excitation occurrence.

\n
The further crucial step is to give an approximate expression for $f_{j_1 0}^{(1)}$ and $f_{0 j_2}^{(1)}$. From \eqref{solf1}, \eqref{G} and \eqref{K}, for  $f_{j_1 0}^{(1)}$ one has 
\be\label{f10}
f_{j_1 0}^{(1)}(\bo R)=\f{M}{2 \pi \hbar^2} \int\!\! d\bo y \,   f_{0 0}^{(0)} (\bo y + \bo a_1)  V_{j_1 0}(\bo y ) \, 
\f{e^{ i k'|\bo R -\bo a_1 - \bo y|}}{|\bo R- \bo a_1 - \bo y|} \, \,,\;\;\;\;\;\; j_1 \neq 0
\ee
and analogously for $f_{0 j_2}^{(1)}$. In order to find the required approximate  expression Mott introduces the following assumptions:

\n
a) the "observation point" $\bo R$ is far away from the origin and the atom, i.e. $|\bo{a_1}|\ll |\bo R|$;

\n
b) the collisions for the $\alpha$-particle are almost elastic,  i.e. $k - k' \ll k$;

\n
c) the $\alpha$-particle has a high momentum $k$. , i.e. $k^{-1}$ is much less than the effective linear dimension of the atom.

\n
Exploiting assumption a) one obtains the asymptotic formula
\be\label{asf1}
f_{j_1 0}^{(1)}(\bo R) \simeq \f{e^{ i k'|\bo R -\bo a_1 |}}{|\bo R- \bo a_1 |} \, \f{M}{2 \pi \hbar^2}  \int\!\! d\bo y \,   f_{0 0}^{(0)} (\bo y + \bo a_1)  V_{j_1 0}(\bo y ) \, 
e^{-i k' \bo u_1(\bo R) \cdot \bo y}
\ee
where
\be\label{u_1}
\bo u_1 (\bo R) = \f{\bo R - \bo a_1}{|\bo R - \bo a_1|}
\ee
Using the explicit expression of $f_{00}^{(0)}$ (see \eqref{00F}) and assumption b)  one can write
\ba\label{asf12}
&&f_{j_1 0}^{(1)}(\bo R) \simeq   \f{e^{ i k'|\bo R -\bo a_1 |}}{|\bo R- \bo a_1 |} \; \mathcal I (\bo u_1 (\bo R)) \\
&& \mathcal I (\bo u_1 (\bo R))= \f{M}{2 \pi \hbar^2}  \int\!\! d\bo y \,  \f{V_{j_1 0}(\bo y )}{|\bo y + \bo a_1 |} \, 
e^{ik \left(  |\bo y + \bo a_1|  - \bo u_1(\bo R) \cdot \bo y \right)} \label{asf13}
\ea
One sees that $f_{j_1 0}^{(1)}(\bo R) $ has the form of a wave diverging from $\bo a_1$, whose amplitude $\mathcal I$ is given by the integral in \eqref{asf13} and it is explicitly dependent on the direction $\bo u_1 (\bo R)$.  
The crucial point is now to evaluate such amplitude.  

\n
By assumptions c),  the integral in \eqref{asf13} is a highly oscillatory integral and then stationary phase arguments can be used.  The leading term of the asymptotic expansion for $k \rightarrow \infty$ is determined by the value of the integrand at the critical points of the phase, i.e. for points $\bo y$ such that 
\be
\nabla_{\bo y} \Big( |\bo y + \bo a_1|  - \bo u_1(\bo R) \cdot \bo y \Big)= \f{\bo y + \bo a_1}{|\bo y + \bo a_1|} - \bo u_1 (\bo R) =0
\ee
On the other hand, the integrand in \eqref{asf13} is very small except in a neighborhood of $\bo y =0$. Therefore one obtains the condition
\be\label{staz}
\bo u_1 (\bo R) \simeq \f{\bo a_1}{|\bo a_1|}
\ee
Using condition \eqref{staz} in \eqref{u_1} one has that the amplitude $\mathcal I$ is significantly different from zero only for those $\bo R$ such that $\bo R - \bo a_1$ is (almost) parallel to $\bo a_1$, i.e. the observation point $\bo R$ must be (almost) aligned with the first atom and the origin. 

\n
From the above argument one concludes that $f_{j_1 0}^{(1)}(\bo R)$ is approximately given by a wave diverging from $\bo a_1$ with an amplitude very small except for $\bo R$ satisfying  \eqref{staz}, i.e. except in a small cone with vertex in $\bo a_1$ and pointing away from the origin.

\n
An analogous  analysis is valid for  $f_{0 j_2}^{(1)}(\bo R)$ and therefore  the computation of the first order term  $F^{(1)}$ is completed.

\n
\label{2 ord} The next step is to consider the second order term $F^{(2)}$. Proceeding as above, one has
\be\label{exF2}
F^{(2)} (\bo{R},\bo{r_1},\bo{r_2}) = \sum_{i_1, i_2} f^{(2)}_{i_1 i_2}(\bo{R}) \Psi^I_{i_1} (\bo{r_1}) \Psi^{II}_{i_2} (\bo{r_2}) 
\ee
and
\ba \label{eqf22}
&& \Big(\!\! -\f{\hbar^2}{2M} \Delta_R +E_{j_1} + E_{j_2} -E \Big)  f^{(2)}_{j_1 j_2}(\bo{R}) \nonumber\\
&&=  \int\!\! d\bo r_1 d \bo r_2 \, \Big(\f{2 e^2}{|\bo R - \bo r_1|} + \f{2 e^2}{|\bo R - \bo r_2|} \Big)  F^{(1)}( \bo R, \bo r_1, \bo r_2) \Psi^I_{j_1}(\bo r_1) \Psi^{II}_{j_2} (\bo r_2) \nonumber\\
&&= \delta_{0 j_2} \sum_{i_1} f^{(1)}_{i_1 0} (\bo R) V_{j_1 i_1}(\bo R - \bo a_1) + f^{(1)}_{j_1 0} (\bo R) V_{0 j_2} (\bo R - \bo a_2)
+   \delta_{ j_1 0} \sum_{i_2} f^{(1)}_{0 i_2} (\bo R) V_{j_2 i_2}(\bo R - \bo a_2) \nonumber\\
&&+ f^{(1)}_{0 j_2} (\bo R) V_{j_1 0} (\bo R - \bo a_1)
\ea
In the case $j_1 , j_2 \neq 0$ one finds
\be\label{eqf23}
 \Big(\!\! -\f{\hbar^2}{2M} \Delta_R +E_{j_1} + E_{j_2} -E \Big)  f^{(2)}_{j_1 j_2}(\bo{R}) =  f^{(1)}_{j_1 0} (\bo R) V_{0 j_2} (\bo R - \bo a_2)+ f^{(1)}_{0 j_2} (\bo R) V_{j_1 0} (\bo R - \bo a_1)
\ee
We recall that $|\bo a_1|<|\bo a_2|$,  $V_{j_1 0} (\bo R - \bo a_1)$ is negligible except for $\bo R \simeq \bo a_1$ and $f^{(1)}_{0 j_2} (\bo R)$ is negligible except for $\bo R$ in a small cone with vertex in $\bo a_2$,  pointing away from the origin. This means that  the last term in the r.h.s. of \eqref{eqf23} is very small. The same kind of argument shows that the first term in the r.h.s. of \eqref{eqf23} is negligible except when the second atom is (approximately) aligned with the first atom and the origin. Thus the main result of the paper follows: 

\n
the probability that both atoms are excited
\be
\int\!\! d\bo R\, |f_{j_1 j_2}^{(2)} (\bo R)|^2 \,, \;\;\;\;\;\;\;\; j_1 , j_2 \neq 0
\ee
is (approximately) zero unless $\bo a_1$, $\bo a_2$ and the origin lie (approximately) on the same straight  line. If one agrees that the (amplified) effect of the excitations of the atoms is the true observed phenomenon in a cloud chamber then the result can be rephrased saying that one can only observe straight tracks. In this sense Mott provides an explanation of the straight tracks observed in the chamber entirely based on the Schr\"odinger equation.


\vs

\section{Concluding remarks}
\vs

\n
We want to conclude with few comments on the different theoretical approaches to the cloud chamber problem discussed in the previous sections. 

\n
As it was pointed out before, according to  Born and Heisenberg, it is definitely equivalent to consider the atom of the vapor as a (classical) measurement device (case a)) or as a part of the quantum system to be described by Schr\"odinger dynamics (case b)).

\n
Such a position is meant to guarantee  the consistence of the standard interpretation of quantum theory.  In particular, the authors are interested in stressing the  unavoidable role of wave packet reduction as  the crucial rule ensuring the right correspondence between theory and observed physical world. From a purely foundational point of view their reasoning aims to make  the axiomatic scheme work in any chosen way to address the cloud chamber dynamical problem and, as such,  it has been adopted and shared by the majority of the physics community for a long time.

\n
However, from the concrete point of view of the physical description of  quantum systems, the equivalence of the two approaches a) and b)  seems difficult to be maintained. 
In fact, one may find hard to accept the claim  that an atom of the vapor is a classical measurement device of the position of the $\alpha$-particle. After all, the atom is a microscopic system on the same ground of the $\alpha$-particle and there is no a priori reason to regard it as a classical system. Agreeing with this point of view, one should concede that approach  b) is surely more natural. In particular, it has the important advantage to allow a quantitative analysis  taking into account explicitly the physical parameters characterizing system and  environment. It is  only on the basis of such a quantitative investigation that it is possible to clarify the conditions under which the interaction with the environment produces the appearance  of classical trajectories. 

\n
We want to stress this crucial point: the classical behavior of the $\alpha$-particle is far from being universal.  A completely dissimilar behavior has to be expected if the values of the physical parameters are different.  On the other hand, consider Heisenberg's argument about ``the atom as a measurement apparatus'', within the framework we called approach a). The resulting reduction of the wave packet should drive the $\alpha$ particle in a state where the position is close to the ionized atom and the momentum has a direction deduced by the successive positions around the source (at time zero) and around the ionized atom (after the flight time to the first ionized atom). The argument seems independent of the interaction strength, while quantum analysis clarifies that, as expected,  only an almost zero-angle scattering process  (very low energy exchange) can guarantee that particle momentum will approximately lay in a small cone with axis on the line connecting source and ionized atom.

\n
A further point to be emphasized concerns the relevance of the result obtained  in the approach b). It is shown that, under the right conditions and with high probability, one can only observe straight tracks. This is a highly non trivial result  obtained exploiting Schr\"odinger dynamics without having any recourse to the wave packet reduction rule.  The reduction should only be invoked at the stage of the "observation of the actual track" described by the $\alpha$-particle. It is remarkable that, as we have seen in section 3,  this aspect was  understood by Darwin even before the quantitative analysis performed by Mott.

\n
On the basis of the above considerations, it seems to us  that, from the physical point of view,  the approach followed by  Mott is more significant and challenging. In some sense, one can say that it establishes the validity conditions  of the effective dynamical behavior of the system which is only postulated in case a) without real physical motivations. 
It is interesting to notice that  awareness of this fact is clearly expressed by Darwin while it remains only implicit in Mott, who  prudently avoids expressing a preference.

\n
From an historical point of view, it is interesting to understand the reasons why the line of research initiated by Mott was not further developed and remained almost neglected for many years. One reason could be the influence and the authority of the position expressed by Born and Heisenberg. The consequence has been  to discourage  the new approach to the problem  with the motivation that it was ineffectively more complicated without giving real advantages from the conceptual point of view.

\n
The whole problem  of  classical behavior in quantum systems has been rediscovered, starting around the eighties of last century, when   remarkable experimental progresses have made possible a detailed  exploration of the classical/quantum border. 
This progress has motivated the development of decoherence theory,  based on the construction of theoretical  models of "system plus environment" where the dynamics of  the  emergence of classical behavior in a quantum system could be analyzed and quantified. To this aim, any investigation based on the  wave packet reduction is of no help and one is forced to consider the problem entirely in the context of Schr\"{o}dinger dynamics.

\n
As examples of successful applications of decoherence theory we mention  the reduction of  interference effects for the state of a heavy particle due to scattering by light particles  (\cite{jz}, \cite{hs}, \cite{afft}) or  the molecular localization for  pyramidal molecules, like $NH_3$, due to the dipole-dipole interaction among the molecules (\cite{cj}, \cite{gms}). 

\n
It is worth underlining that the pioneering work of Mott is entirely within the same  line of thought.  Moreover, we emphasize that the relevance of such a research is more related to the concrete and applied  analysis of quantum systems rather than to foundational aspects.

\n
We want to conclude commenting on some limitations present in Mott's analysis and on possible developments of his work.

\n
The first point concerns  the use of the stationary Schr\"odinger equation which prevents a clear description of the time evolution of the three-particle system. In particular, it is missing a physically meaningful definition of the initial state and an explicit time-dependent analysis of the successive interactions of the $\alpha$-particle with the two atoms. 
A second aspect is related to the proof techniques used by Mott which exploit  perturbation theory and stationary phase arguments. There is a clear lack of control of the applicability conditions of these methods and there is no estimate of the error resulting the various approximations used by the author (for a recent attempt to revisit Mott's model we refer to \cite{dft2}).

\n
We finally notice that a further development of Mott's work is lacking in the literature. 
In particular, the model should be generalized  considering a more generic initial state for the particle, the presence of external force fields and an environment made by an arbitrary number of model atoms. 
In this way one could analyze the possible emergence of a classical trajectory of the quantum particle in a more realistic model. In our opinion, such  analysis would surely be  of great interest  both from the theoretical and from the applicative point of view.

\vs
\vs
\vs
\vs
\vs

\vs
\vs
\vs


\begin{thebibliography}{99}

\bibitem[AFFT]{afft} Adami R., Figari R., Finco D., Teta A., On the asymptotic dynamics of a quantum system composed by heavy and light particles.  {\em Comm.   Math. Phys.} {\bf 268}, no. 3, 819-852, 2006.


\bibitem[BV]{bv} Bacciagaluppi G., Valentini A., {\em Quantum Theory at the Crossroads: Reconsidering the 1927 Solvay Conference}, Cambridge University Press, 2009.

\bibitem[Boh]{boh} Bohr N., The quantum postulate and the recent development of atomic theory. {\em Nature}, {\bf 121}, 580-590, 1928.

\bibitem[Bor]{bor} Born M., 
Zur Quantenmechanik der Stossvorg\"{a}nge. {\em  
Z. Phys.},  {\bf 37}, 863-867, 1926. Eng. trans. reprinted in: Wheeler J.A., Zurek W., {\em Quantum Theory and Measurement}, Princeton University Press, 1983. 

\bibitem[BJ]{bj} Born M., Jordan  P.,  
Zur Quantenmechanik. 
{\em Z. Phys.},  {\bf 34}, 858, 1925. Eng. trans. reprinted in: van der Waerden B.L., {\em Source of Quantum Mechanics}, Dover Publications, Inc., 1967.



\bibitem[CK]{ck} Carazza B., Kragh H., Classical behavior of macroscopic bodies from quantum principles: early discussions. {\em Arch. Hist. Exact Sci.}, {\bf 55}, 43-56, 2000.

\bibitem[CJ]{cj} Claverie P., Jona-Lasinio G., Instability of tunnelling and the concept of molecular structure in quantum mechanics: the case of pyramidal molecules and the enantiomer problem. {\em Phys. Rev. A}, {\bf 33}, 2245-2253 (1986).




\bibitem[Cu]{c} Cushing J.T., {\em Quantum Mechanics, Historical Contingency and the Copenhagen Hegemony}, The University of Chicago Press, 
1994.

\bibitem[Da]{da} Darwin C.G., A collision problem in the wave mechanics. {\em Proc. R. Soc.
Lond. A}, {\bf  124}, 375-394, 1929.

\bibitem[DFT1]{dft1} Dell'Antonio G., Figari R., Teta A., Joint excitation probability for two harmonic oscillators in dimension one and the Mott problem.  {\em J.  Math. Phys.} {\bf 49}, n. 4,  042105, 2008.

\bibitem[DFT2]{dft2} Dell'Antonio G., Figari R., Teta A., A time dependent perturbative analysis for  a quantum particle in a cloud chamber. {\em Ann.  H.  Poincare'} {\bf 11}, no. 3, 539-564, 2010.



\bibitem[Ga]{ga} Gamow G., Zur Quantentheorie des Atomkernes.  {\em Z. Phys.}, {\bf 51}, 204, 1928.

\bibitem[GJKKSZ]{gjkksz} Giulini D., Joos E., Kiefer C., Kupsch J., Stamatescu I.-O., Zeh
H.D., {\em Decoherence and the Appearance of a Classical World in
Quantum Theory}, Springer, 1996.

\bibitem[GMS]{gms} Grecchi V., Martinez A., Sacchetti A., Destruction of the beating effect for a non-linear Schr\"{o}dinger equation. {\em Comm. Math. Phys.}, {\bf 227}, 191-209, 2002.




\bibitem[He1]{he1} Heisenberg W.,  
\"{U}ber quantentheoretische Umdeutung kinematischer und mechanischer Beziehungen. {\em Z. Phys.},  {\bf 33}, 879-893, 1925. Eng. trans. reprinted in:  van der Waerden B.L., {\em Source of Quantum Mechanics}, Dover Publications, Inc., 1967.

\bibitem[He2]{he2} Heisenberg W., \"{U}ber den anschaulichen Inhalt der quantentheoretischen Kinematik und Mechanik.  {\em Z. Phys.}, {\bf 43}, 172-198, 1927. Eng. trans. reprinted in: Wheeler J.A., Zurek W., {\em Quantum Theory and Measurement}, Princeton University Press, 1983. 

\bibitem[He3]{he3} Heisenberg W., {\em The Physical Principles of the Quantum Theory}, The University of Chicago Press, 1930.

\bibitem[HS]{hs} Hornberger K., Sipe J.E., Collisional decoherence reexamined. 
{\em Phys. Rev. A} {\bf 68}, 012105, 1-16, 2003.



\bibitem[Ja]{ja} Jammer M., {\em The Conceptual Development of Quantum Mechanics}, 2nd ed., American Institute of Physics, 
1989.

\bibitem[JZ]{jz} Joos E., Zeh H.D., The emergence of classical properties through
interaction with the environment. {\em Z. Phys.} {\bf B59}, 223-243, 1985.



\bibitem[LR]{lr} Leone M., Robotti N., A note on the Wilson cloud chamber (1912). {\em Eur. J.  Phys.}, {\bf 25}, 781-791, 2004. 

\bibitem[Mo]{mo} Mott  N.F., The wave mechanics of $\alpha$-ray tracks. {\em Proc. R. Soc.
Lond. A}, {\bf  126}, 79-84, 1929. Reprinted in: Wheeler J.A., Zurek W., {\em Quantum Theory and Measurement}, Princeton University Press, 1983. 

\bibitem[Ro]{ro} Robert D.,  Semi-classical approximation in quantum mechanics. A survey of old and recent mathematical results. {\em Helv. Phys. Acta}, {\bf 71}, 44-116, 1998.


\bibitem[Sc]{sc} Schr\"odinger E., {\em Collected papers on Wave Mechanics}, Chelsea Pub. Co, 2nd edition, 1978.

\bibitem[St]{st} Stepansky B.K.,  Ambiguity: aspects of the wave-particle duality. {\em Brit. J. Hist. Sci.}, {\bf 30}, 375-385, 1997.



\bibitem[vo]{vo} von Neumann J., {\em Mathematische Grundlagen der Quantenmechanik}, Springer-Verlag, 1932. Engl. trans. {\em Mathematical Foundations of Quantum Mechanics}, Princeton University Press, 1955.

\end{thebibliography}
\end{document}